\documentclass[twocolumn, 10pt]{article}

\usepackage[utf8]{inputenc}

\usepackage[pdftex]{graphicx}
\graphicspath{{figures/}}
\DeclareGraphicsExtensions{.pdf,.jpeg,.png}
\usepackage{epstopdf}

\usepackage[cmex10]{amsmath}
\usepackage{amsfonts}
\usepackage{amsthm}

\usepackage{algorithmic}
\usepackage[caption=false,font=footnotesize]{subfig}
\usepackage{ctable}

\usepackage{authblk}

\theoremstyle{plain}
\newtheorem{assumption}{Assumption}

\renewenvironment{abstract}{%
\hfill\begin{minipage}{0.99\textwidth}
\rule{\textwidth}{1pt}}
{\par\noindent\rule{\textwidth}{1pt}\end{minipage}\vspace{10mm}}

\makeatletter
\renewcommand\@maketitle{%
\hfill
\begin{minipage}{0.95\textwidth}
\vskip 2em
\let\footnote\thanks 
{\LARGE \textbf \@title \par }
\vskip 1.5em
{\large \@author \par}
\end{minipage}
\vskip 1em \par
}
\makeatother

\begin{document}

\title{On the Scalability of LISP Mappings Caches}

\author{Florin Coras}
\author{Jordi Domingo-Pascual}
\author{Albert Cabellos-Aparicio}
\affil{\small Universitat Polit\'ecnica de Catalunya (BarcelonaTECH), Barcelona, Spain}

\date{}

\twocolumn [
\maketitle
\begin{@twocolumnfalse}
\begin{abstract} 
    \bf
    The Locator/ID Separation Protocol (LISP) limits the growth of the
    Default-Free Zone routing tables by creating a highly aggregatable and
    quasi-static Internet core. However, LISP pushes the forwarding state to
    edge routers whose timely operation relies on caching of location to
    identity bindings. In this paper we develop an analytical model to
    study the asymptotic scalability of the LISP cache. Under the assumptions
    that (i) long-term popularity can be modeled as a constant Generalized
    Zipf distribution and (ii) temporal locality is predominantly determined
    by long-term popularity, we find that the scalability of the LISP cache is
    O(1) with respect to the amount of prefixes (Internet growth) and users
    (growth of the LISP site). We validate the model and discuss the accuracy
    of our assumptions using several one-day-long packet traces.

\end{abstract}
\end{@twocolumnfalse}
]

\section{Introduction}\label{sec:introduction}
The growth of the Default-Free Zone (DFZ) routing
tables~\cite{rfc4984} and associated churn observed in recent years has led to
much debate as to whether the current Internet infrastructure is
architecturally unable to scale. Sources of the problem were found to be
partly organic, generated by the ongoing growth of the topology, but also
related to operational practices which seemed to be the main drivers behind
prefix deaggregation within the Internet's core. Diverging opinions as to how the latter
could be solved triggered a significant amount of research that finally materialized in
several competing solutions (see~\cite{rfc6115} and the references therein).

In this paper we focus on location/identity separation type of approaches in
general, and  consider the Locator/ID Separation Protocol
(LISP)~\cite{saucez:lisp} as their particular instantiation. LISP semantically
decouples identity from location, currently overloaded by IP addresses, by
creating two separate namespaces that unambiguously address end-hosts
(identifiers) and their Internet attachment points (locators). This new
indirection level has the advantage that it supports the implementation of
complex traffic engineering mechanisms but at the same time enables the
locator space to remain quasi-static and highly
aggregatable~\cite{rfc7215}.

Although generally accepted that location/identity type of solutions alleviate the
scalability limitations of the DFZ, they also push part of the forwarding
complexity to the edge domains. On the one hand, they require mechanisms to
register, distribute and retrieve bindings that link elements of the two new
namespaces. On the other, LISP routers must store in use mappings to speed-up
packet forwarding and to avoid generating floods of resolution requests. 
This then begs the question: \emph{does the newly introduced LISP edge cache
scale?}

This paper provides an analytical answer by analyzing the scalability of the
LISP cache with respect to the growth of the Internet and growth of the LISP
site. To this end we leverage the working-set theory~\cite{denning:ws_model}
and previous results that characterize temporal locality of reference
strings~\cite{breslau:web_and_zipf, jin:web_tloc} to develop a model that
relates the LISP cache size with the miss-rate. We find that the relation
between cache-size and miss-rate only depends on the popularity distribution
of destination prefixes. Additionally, for a given miss rate, as long as the
popularity follows a Generalized-Zipf distribution, the LISP cache size scales
constantly O(1) with respect to the growth of the Internet and the number
users, if the last two do not influence the popularity distribution. If this
does not hold then the cache scales linearly O(N). To support our results, we
also analyze the popularity distribution of destination prefixes in several one day
real-world packet traces, from two different networks and spanning a period of
$3.5$ years.

The rest of the paper is structured as follows. We provide a brief overview of
LISP in Section~\ref{sec:background}. In Section~\ref{sec:cache_model} we
derive the cache model under a set of assumptions and thereafter discuss its
predictions and implications for LISP.  In Section~\ref{sec:evaluation}
we present empirical evidence that supports our assumptions and evaluate the
model, while in Section~\ref{sec:rw} we discuss the related work. Finally, we
conclude the paper in Section~\ref{sec:conclusions}.
 
\section{LISP Background}\label{sec:background}
LISP~\cite{saucez:lisp} belongs to the family of proposals that implement
a location/identity split in order to address the scalability concerns of the
current Internet architecture. The protocol specification has recently
undergone IETF standardization~\cite{rfc6830}, however development and
deployment efforts are still ongoing. They are supported by a sizable
community spanning both academia and industry and rely for testing on a large
experimental network, the LISP-beta network~\cite{lisp:testbed}. 

The goal of splitting location and identity is to insulate core network
routing that should ideally only be aware of location information (locators),
from the dynamics of edge networks, which should be concerned with the delivery of
information based on identity (identifiers). To facilitate the transition from the current
infrastructure, LISP numbers both namespaces using the existing IP addressing
scheme, thus ensuring that routing within both core and stub networks stays unaltered.
However, as locators and identifiers bear relevance only within their
respective namespaces, a form of conversion from one to the other must be
performed. LISP makes use of encapsulation~\cite{rfc1955} and a directory
service to perform such translation.

\begin{figure}[t]
    \centering
    \includegraphics[width=80mm,keepaspectratio=true]{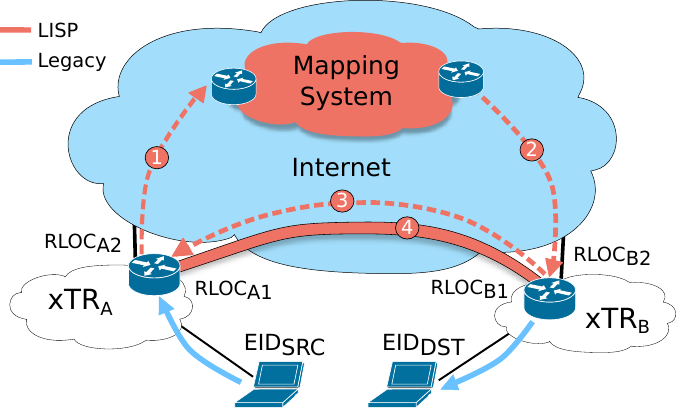}
    \caption{Example packet exchange between $EID_{SRC}$ and $EID_{DST}$ with
    LISP. Following intra-domain routing, packets reach $xTR_{A}$ which
    obtains a mapping binding $EID_{DST}$ to $RLOC_{B1}$ and $RLOC_{B2}$ from the mapping-system
    (steps 1-3). From the mapping, $xTR_{A}$ chooses $RLOC_{B1}$ as destination
    and then forwards towards it the encapsulated packets over the Internet's
    core (step 4). $xTR_{B}$ decapsulates the packets and forwards them to their
    intended destination. }
    \label{fig:lisp-arch}
\end{figure}

Prior to forwarding a host generated packet, a LISP router maps the
destination address, or Endpoint IDentifier (EID), to a corresponding
destination Routing LOCator (RLOC) by means of a LISP specific mapping
system~\cite{draft:lisp-ddt,jakab:lisp-tree}.  Once a mapping
is obtained, the border router tunnels the packet from source edge to
corresponding destination edge network by means of an encapsulation with a
LISP-UDP-IP header. The outer IP header addresses are the RLOCs pertaining to
the corresponding border routers (see Fig.~\ref{fig:lisp-arch}). At the
receiving router, the packet is decapsulated and forwarded to its intended
destination. In LISP parlance, the source router, that performs the
encapsulation, is called an Ingress Tunnel Router (ITR) whereas the one
performing the decapsulation is named the Egress Tunnel Router (ETR). One that
performs both functions is referred to as an xTR.

Since the packet throughput of an ITR is highly dependent on the time needed
to obtain a mapping, but also to avoid overloading the mapping-system, ITRs
are provisioned with map-caches that store recently used EID-prefix-to-RLOC
mappings. Stale entries are avoided with the help of timeouts, called
\emph{time to live} (TTL), that mappings carry as attributes. Whereas,
consistency is ensured by proactive LISP mechanisms through which the xTR
owner of an updated mapping informs its peers of the change.  Intuitively, the
map-cache is most efficient in situations when destination EIDs present high
temporal and/or spatial locality and its size depends on the diversity of the
visited destinations. As a result, performance depends entirely on map-cache
provisioned size, traffic characteristics and the eviction policy set in
place.

\section{Cache Model}\label{sec:cache_model}
We start this section by discussing some of the fundamental properties of network traffic
that may be exploited to gain a better understanding of cache performance. Then,
assuming these properties are characteristic to real network traces we devise a
cache model. Finally we analyze and discuss the predictions of the model. 

\subsection{Sources of Temporal Locality in Network Traffic}\label{sec:temp_locality}

We consider the following formalization of traffic, either at Web page or
packet level, throughout the rest of the paper. Let $D$ be a set of objects
(Web pages, destination IP-prefix, program page etc.). Then, we consider 
traffic to be a strings of references $r_1, r_2, \dots, r_i\dots$ where
$r_i = o \in D$ is a reference at the $i$th unit of time that 
has as destination, or requests, object $o$. Generally, we consider the length
of the reference string to be $N$. Also, note that we use object
and destination interchangeably. 

Two of the defining properties of reference strings, important in
characterizing cache performance, are the heavy tailed \emph{popularity
distribution} of destinations and the \emph{temporal locality} exhibited by
the requests pattern. We discuss both in what follows.  

\subsubsection{Popularity Distribution} copious amounts of studies in fields
varied as linguistics~\cite{zipf:principle_least_effort, montemurro:gzipf},
Web traffic~\cite{breslau:web_and_zipf,mahanti:web_proxy_hierarchy},
video-on-demand~\cite{cha:itube}, p2p overlays~\cite{dan:plaw} and flow level
traffic~\cite{sarrar:leverage_zipf} found the probability distribution of
objects to have a positive skew. Generally, such distributions are coined
Zipf-like, i.e., they follow a power law; whereby the probability
of reference is inversely proportional to the rank of an object. Generally,
the relation is surmised as: $\nu(k) = \dfrac{\Omega}{k^\alpha}$
\noindent where $\nu$ is the frequency, or number of requests observed for an
object, $k$ is the rank, $\Omega=1/H(n,\alpha)$ is a normalizing constant and
$H(n,\alpha)$ is the $n^{th}$ generalized harmonic number. 

It is interesting to note that although Zipf's law has its origins in
linguistics, it was found to be a poor fit for the statistical behavior of
words frequencies with low or mid-to-high values of the rank variable. That
is, it does not fit the head and tail of the distribution.
Furthermore, it's extension due to Mandelbrot (often called the
Zipf-Mandelbrot law) only improves the fitting for the head of the
distribution.
Such discrepancies were also observed for Web based and p2p reference strings.
Often the head of the distribution is flattened, i.e., frequency is less than the
one predicted by the law, or the tail has an exponential cutoff or a faster
power law decay~\cite{montemurro:gzipf, dan:plaw}. But these differences are usually 
dismissed on the basis of poor statistics in the high ranks region
corresponding to objects with a very low frequency. 

Nevertheless, Montemurro solved recently the problem in linguistics by
extending the Zipf-Mandelbrot law such that for high ranks the tail
undergoes a crossover to an exponential or larger exponent power-law decay.
Surprisingly, he found this features, i.e. deviations from the Zipf-like
behavior, to hold especially well when very large
corpora~\cite{montemurro:gzipf} are considered. We further refer to this model
as the Generalized Zipf law or GZipf and, in light of these observations, we
assume the following:

\begin{assumption}\label{prop:gzipf}
    The popularity distribution of destination IP-prefix reference strings can be
    approximated by a GZipf distribution.
\end{assumption}

\subsubsection{Temporal locality} can be informally defined as the property that a recently
referenced object has an increased probability of being re-referenced. 
One of the well established ways of measuring the degree of locality of
reference strings is the inter-reference distance distribution.

Breslau et al. found in ~\cite{breslau:web_and_zipf} that strings generated
according to the Independent Reference Model (IRM), that is, assuming that
references are independent and identically distributed random variables, from
a popularity distribution have an inter-reference distribution similar to that
of the original string. Additionally, they inferred that the probability of an
object being re-referenced after $t$ units of time is proportional to $1/t$.
Later, Jin and Bestavros proved that in fact temporal locality emerges from
both long-term popularity and short-term correlations. However, they found
that the inter-reference distance distribution is mainly induced through
long-term popularity and therefore is insensitive to the latter. Additionally,
they showed that by ignoring temporal correlations and assuming a Zipf-like
popularity distribution, an object's re-reference probability after $t$ units
of time is proportional to  $1/t^{(2-1/\alpha)}$. These observations then lead
to our second assumption:

\begin{assumption}\label{prop:tloc}
    Temporal locality in destination IP-prefix reference strings is mainly due
    to the prefix popularity distribution.
\end{assumption}

We contrast the two assumptions with the properties of several packet-level
traces in~\ref{sec:evaluation}. In what follows we are interested in
characterizing the inter-reference distribution of a GZipf distribution and
further on the cache miss rate using the two statements as support.

\subsection{GZipf generated inter-reference distribution}

In this section we compute the inter-reference distance distribution for a GZipf
popularity. The result is an extension of the one due to Jin and Bestavros for
a Zipf-like popularity. As a first step we compute the inter-reference
distribution for a single object and then by integration obtain the average
for the whole reference string, which we denote by $f(t)$.

If $\nu$ is the normalized frequency, namely, the number of reference to an
object divided by the length of the reference string $N$, then, as shown
in~\cite{montemurro:gzipf} the probability of observing objects with frequency
$\nu$ in the reference string is:

\begin{equation}
    p_{\nu}(\nu)\propto\dfrac{1}{\mu\nu^r+(\lambda-\mu)\nu^q}
    \label{eq:gzipf_pdf}
\end{equation}

\noindent where $1 \le r<q$ are the exponents that control the slope of the
power laws in the two regimes and $\mu$ and $\lambda$ are two constants that
control the frequency for which the tail undergoes the crossover. 

From Assumption~\ref{prop:tloc} it follows that references to an object are
independent whereby the inter-reference distance $t$ is distributed exponentially
with expected value of $1/\nu$. Then, if we denote by $d(t, \nu)$ the
number of times the inter-reference distance for an object with frequency
$\nu$ is $t$, we can write:

\begin{equation}
    d(t, \nu) \sim (\nu N-1) \nu e^{-\nu t} 
    \label{eq:dtf}
\end{equation}

If $\nu_{min}$ and $\nu_{max}$ are the minimum and respectively the
maximum normalized frequency observed for the reference string, we can compute
the inter-reference distance for the whole string as:

\begin{eqnarray}
    f(t) &\sim& \int_{\nu_{min}}^{\nu_{max}} p_{\nu}(\nu)\,d(t,\nu) \mathrm{d}\nu \nonumber \\
    &=& \int_0^1 \dfrac{(\nu N-1)\nu e^{-\nu t} }{\mu\nu^r+(\lambda-\mu)\nu^q} \mathrm{d}\nu
    \label{eq:ir}
\end{eqnarray}

Unfortunately, the integral is unsolvable, nevertheless, we can still characterize the
properties of $f(t)$ in the two regimes of the GZipf
distribution. In the high frequency region, where term having $q$ as exponent dominates the
denominator we can write:

\begin{eqnarray}
    f_q(t)  &\sim& \int_{\nu_k}^1 \dfrac{\nu^2\ e^{-\nu t} }{\nu^q}
    \mathrm{d}\nu \nonumber \\
                        &=& \dfrac{\Gamma(3-q, \nu_k t)}{t^{3-q}} 
    \label{eq:ir_q}
\end{eqnarray}

\noindent where, $\Gamma(n,z) =\int_z^\infty x^{n-1} e^{-x}\mathrm{d}x$ is the
incomplete Gamma function.  $\nu_k=(\mu/(\lambda-\mu))^{1/(q-r)}$ is the
frequency for which the two terms that make up the denominator are equal. It
is useful to note that for low $t$ values that correspond to high frequencies
the  nominator presents a constant plateau that quickly decreases, or bends,
at the edge as $t\to 1/\nu_{k}$. Therefore, we can approximate:

\begin{equation}
    f_q(t)\sim \dfrac{1}{t^{3-q}}
    \label{eq:fq_asym}
\end{equation}

Similarly, it may be shown that for low frequencies, that is, in the region
where term with $r$ as exponent dominates:

\begin{equation}
    f_r(t)\sim \dfrac{1}{t^{3-r}}
    \label{eq:fr_asym}
\end{equation}

Finally, we conclude that the inter-reference distance distribution can be
approximated by a piece-wise power-law. Our result is similar to the single
sloped power-law obtained by Jin under the assumption of Zipf distributed
popularity or the empirical observations by Breslau et. al
in~\cite{breslau:web_and_zipf} for Web reference strings. However, due to its
general form it should be able to capture the properties of more varied
workloads. 
In the following section we use the inter-reference distance
distribution together with the working-set theory to deduce the miss rate of
an LRU cache.

\subsection{A Cache Model} 

Denning proposed the use of the working-set as a tool to capture the set of
pages a program must store (cache) in memory such that it may operate at a
desired level of efficiency~\cite{denning:ws_model}. The idea is to estimate a
program's locality, or in-use pages, with the help of a sliding window of
variable length looking into the past of the reference string.  In their
seminal work characterizing the properties of the
working-set~\cite{denning:ws_properties}, Denning and Schwartz showed that the
average inter-reference distance is the slope of the average miss rate,
which at its turn is the slope of the average working-set size, both taken as
functions of the window size. The result is of particular interest as it
provides a straightforward link between the properties of the reference string
and the performance of a cache that uses the least recently used (LRU)
eviction policy but whose size varies. To understand the latter consider that
the size of the working-set for a given window depends on the number of
unique destinations within the window, which may vary. Still, under
the condition 
that the reference string is obtained with IRM, the working-set size will be
normally distributed with a low variance. We can approximate it as being
constant and as a result the cache modeled by the working-set becomes an LRU
of fixed size. 

We leverage in what follows the result above to deduce miss rate of an LRU
cache when fed by a reference string obtained using IRM and a GZipf popularity
distribution. The miss rate for the upper part of $f(t)$ is:

\begin{equation}
    m_q(t) = - \int \dfrac{C}{t^{3-q}} \mathrm{d}t = - C \dfrac{t^{q-2}}{q-2}
    \label{eq:mtq}
\end{equation}

\noindent where, $t < 1/\nu_k$, $1<q<2$ and $C$ is a normalizing constant
which ensures that $\sum\limits_{t=1}^{N-1} C f(t) = 1$. We can further
compute the average working-set size as:

\begin{equation}
    s_q(t) =  \int C\dfrac{t^{q-2}}{q-2} \mathrm{d}t = -C
    \dfrac{t^{q-1}}{(q-1)(q-2)}
    \label{eq:stq}
\end{equation}

To obtain the miss rate as a function of the cache size, not of the
inter-reference distance, we take the inverse of $s_q$ and replace it in
(\ref{eq:mtq}). For $s < s_q(1/\nu_k)$ we get:

\begin{eqnarray}
    m_q(s) &=& C^{\dfrac{1}{q-1}} (2-q)^{-\dfrac{1}{q-1}}
    (q-1)^{\dfrac{q-2}{q-1}} s^{\dfrac{q-2}{q-1}} \nonumber \\
            &\propto& s^{1-\dfrac{1}{q-1}}
    \label{eq:msq}
\end{eqnarray}

This suggests that the asymptotic miss rate as a function of cache size is a
power law of the cache size with an exponent dependent on the slope of the
popularity distribution. Similarly, for large inter-reference distances, when
$s>s_r(1/\nu_k)$:

\begin{equation}
     m_r(s) \propto s^{1-\dfrac{1}{r-1}}
    \label{eq:msr}
\end{equation}

Then, for a reference string whose destinations have a GZipf popularity distribution
and where the references to objects are independent, we find that the miss
rate presents two power-law regimes with exponents only dependent on the
exponents of the popularity distribution and the cache size. We test the ability of the
equations to fit empirical observations in~\ref{sec:cache_res}.

\subsection{Cache Performance Analysis}\label{sec:cache_bounds}
We now investigate how cache size varies with respect to the parameters of the
model if the miss rate is held constant. By inverting (\ref{eq:msq}) and (\ref{eq:msr}) we obtain the cache size
as a function of the miss rate:
\begin{equation}
    s(m)= 
    \begin{cases}
        g(q)\,m^{1-\dfrac{1}{2-q}}, & \quad m \le m_k \\
        g(r)\,m^{1-\dfrac{1}{2-r}}, & \quad m > m_k \\
    \end{cases}
    \label{eq:sm_both}
\end{equation}

\noindent with $g(x)=-C^{\frac{1}{2-x}}\dfrac{ (2-x)^{\frac{x-1}{x-2}}
}{2-3x+x^2}$,  $m_k = \dfrac{C}{\nu_k^{r-2} (2-r)}$,
$\nu_k=\left(\dfrac{\mu}{\lambda-\mu}\right)^{q-r}$ and $0<m<1$. 

We see that $s(m)$ is \emph{independent} of both the number of packets $N$
and the number of destinations $D$ and is sensible
only to changes of the slopes of the popularity distribution $q$, $r$ and the
frequency at which the two slopes intersect, $\nu_k$. We do note that $C$ does depend
analytically on $N$ as it can be seen by considering $C$'s defining expression
(see discussion of~(\ref{eq:mtq})):
$1/C=H(1/\nu_k,3-q)-\zeta(3-r,N)+\zeta(3-r, 1/\nu_k)$ where
$H(n,m)=\sum\limits_{k=1}^n 1/k^m$ is the generalized harmonic number of order
$n$ of $m$ and $\zeta(s,a)=\sum\limits_{k=0}^\infty 1/(k+a)^s$ the Hurwitz
Zeta function.  However, the first and last terms of the expression depend
only on popularity parameters while the middle one quickly converges to a
constant as $N$ grows. Whereby it is safe to assume $C$ constant with
respect to $N$ and consequently that the number of packets does not influence
$s(m)$.

\begin{figure}[h]
    \centering
    \includegraphics[width=0.5\textwidth,keepaspectratio=true]{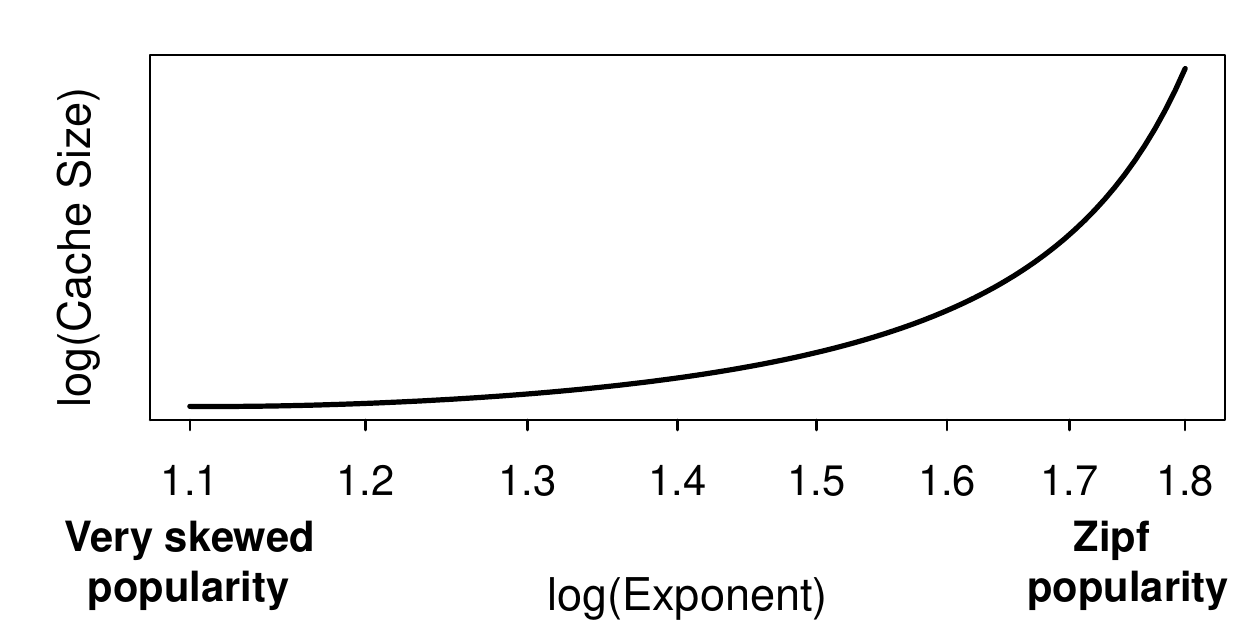}
    \caption{Cache size as a function of a GZipf exponent for a fixed miss rate}
    \label{fig:cs_vs_q}
\end{figure}

On the other hand, if the parameters of the popularity distribution are
modified, some interesting dependencies can be uncovered. For brevity, we explore only the
case when $q$ and $r$ vary but still respect the constraint that $1<r<q<2$. 
When both exponents jointly change, the cache size required to
maintain the miss rate will qualitatively vary as depicted in
Fig.~\ref{fig:cs_vs_q}. Specifically, as their value approach $1$, that is,
when the popularity distribution is strongly skewed, cache size asymptotically
goes to a low value constant, whereas when the exponent approaches $2$, the
required cache size grows very fast, notice the superlinear growth in the
log-log scale. Despite not being indicated by (\ref{eq:sm_both}), $s(m)$ is
defined when $q$ or $r$ are $2$, that is, it does not grow unbounded. The
expression can be obtained if we replace $q$ by $2$ in~(\ref{eq:mtq}) and
recompute all equations:
\begin{equation}
    s(m) = (C+m)\, e^{-\dfrac{m}{C}} 
    \label{sm_log}
\end{equation}

\subsection{Discussion of Asymptotic Cache Performance and Impact}

Using the results of the analysis performed in the previous section we are now
interested to characterize the asymptotic scalability of the LISP cache size
with respect to (i) the number of users in a LISP site (ii) the size of the
EID space and (iii) the parameters of the popularity distribution. To simplify
the discussion, we assume there are no interactions between the first two and
the third:

\begin{assumption}
    The destination prefix popularity distribution is independent of the
    number of users in a LISP site and the size of the EID space.
\end{assumption}

Whereby (i) contemplates the variation of the number of packets, $N$ (ii) the
variation of the number of destinations $D$ and (iii) the variation of the
GZipf parameters $q$, $r$, $\mu$ and $\lambda$, independently. We acknowledge that the
popularity distribution may be influenced by a multitude of factors, and in
particular by the growth of the users generating the reference string.
Nonetheless, we argue that our assumption does make practical sense.  For
instance, a typical LISP router is expected to serve hundreds to thousands of
clients so fluctuations proportional to the size of the user set should
not affect overall homogeneity and popularity distribution.  Additionally,
although user interest in content quickly changes, the same is not necessarily
true for the content sources, i.e., prefixes from where the content is
served, which the user cannot typically select. This split between content and its
location can result in relatively stable popularity distribution of the 
prefixes despite the dynamic popularity of actual content. We show an example
network where this assumption holds in Section~\ref{sec:pop_assumption}.

In the previous section we found that when the parameters of the popularity
distribution are held constant, the cache size is independent of both the
number of packets and destinations. As a result, cache size scales constantly,
O(1), with the number of users within a LISP site and the size of EID-prefix space
for a fixed miss rate. This observation has several fundamental implications
for LISP's deployment. First, caches for LISP networks can be designed and
deployed for a desired performance level which subsequently does not degrade
with the growth of the site and the growth of the Internet address space.
Second, splitting traffic between multiple caches (i.e., routers) for
operational purposes, within a large LISP site, does not affect cache
performance.  Finally, signaling, i.e., the number of Map-Request exchanges,
grows linearly with the number of users if no hierarchies or cascades of
caches are used. This because the number of resolution requests is $m(s)\,N$.

If the previous assumption does not hold, then, in the worst case, the cache
size scales linearly with $|D|$. This follows if we consider that, as the
growth of $N$ and $D$ flatten the distribution, thus leading to a uniform
popularity, the cache size for a desired miss rate becomes proportional to the
$|D|$.

\section{Empirical Evidence of Temporal Locality}\label{sec:evaluation}
In this section we verify the accuracy of our assumptions regarding the
popularity distribution of destination prefixes and the sources of locality in
network traffic. We also verify the accuracy of the predictions regarding the
performance of the LISP cache empirically. But first, we present our datasets
and experimental methodology.

\subsection{Packet Traces and Cache Emulator}

We use four one-day packet traces that only consist of egress traffic for our
experiments. Three were captured at the 2Gbps link that connects our
University's campus network to the Catalan Research Network (CESCA) and span a
period of 3.5 years, from 2009 to 2012. The fourth was captured at the 10Gbps
link connecting CESCA to the Spanish academic network (RedIris) in 2013. UPC
campus has about 36k users consisting generally of students, academic staff
and auxiliary personnel while CESCA provides transit services for 89
institutions that include the public Catalan schools, hospitals and
universities. 
The important properties of the datasets are summarized in
Table~\ref{tab:traces}.

\begin{table*}[t]
    \centering
    \caption{Datasets Statistics}
    \label{tab:traces}
    \begin{tabular}{lcccc}
        \toprule
                & \textbf{upc 2009} & \textbf{upc 2011} & \textbf{upc 2012} & \textbf{cesca 2013} \\ \midrule[0.09em]
        Date    & 2009-05-26 & 2011-10-19 & 2012-11-21 & 2013-01-24\\ \midrule 
        Packets & 6.5B & 4.05B & 5.57B & 20B \\ \midrule
        Av. pkt/s & 75.3k & 46.9k & 64.4k & 232k  \\ \midrule
        Prefixes  & 92.8k & 94.9k & 109.4k & 143.7k \\ \midrule
        Av. pref/s & 2.3k & 1.95k & 2.1k & 2.56k \\         
        \bottomrule
    \end{tabular}
\end{table*}

\begin{table*}[t]
    \centering
    \caption{Routing Tables Statistics}
    \label{tab:rt}
    \begin{tabular}{lcccc}
        \toprule
        & \textbf{upc 2009} & \textbf{upc 2011} & \textbf{upc 2012} & \textbf{cesca 2013} \\ \midrule[0.09em]
        $\texttt{BGP}_{RT}$ & 288k & 400k & 450k & 455k \\ \midrule
        $\texttt{BPG}_{\phi}$ & 142k & 170k & 213k & 216k \\ \midrule
        $\rho$ & 0.65 & 0.55 & 0.51 & 0.66 \\
        \bottomrule
    \end{tabular}
\end{table*}

At the time of this writing there exists no policy as to how EID-prefixes are
to be allocated. However, it is expected and also the practice today in the
LISP-beta network to allocate EIDs in IP-prefix-like blocks. Consequently we
performed our analysis considering EID-prefixes to be of BGP-prefix granularity. For each
packet within a trace we find the associated prefix using BGP routing tables
downloaded form the RouteView archive~\cite{routeviews} that match the trace's
capture date. We filtered out the more specific prefixes from the routing
tables as they are generally used for traffic engieering and LISP offers a
more efficient management of these operational needs. Table~\ref{tab:rt} gives
an overview of the original ($\texttt{BGP}_{RT}$), and filtered
$\texttt{BGP}_{\phi}$ routing table sizes as well as the ratio ($\rho$) between the
filtered routing table size and the the number of prefixes observed within
each trace. Both UPC and CESCA visit daily more than half of the prefixes
within $\texttt{BGP}_{\phi}$.

Apart from the popularity and temporal locality analysis we also implemented
an LISP ITR emulator to estimate LRU map-cache performance using the traces and
the routing tables as input. We compare the predictions of our
cache model with the empirical results in~\ref{sec:cache_res}.

\subsection{Popularity Distribution}\label{sec:pop_assumption}

\begin{figure}[t!]
    \centering
    \includegraphics[width=0.5\textwidth, keepaspectratio=true]{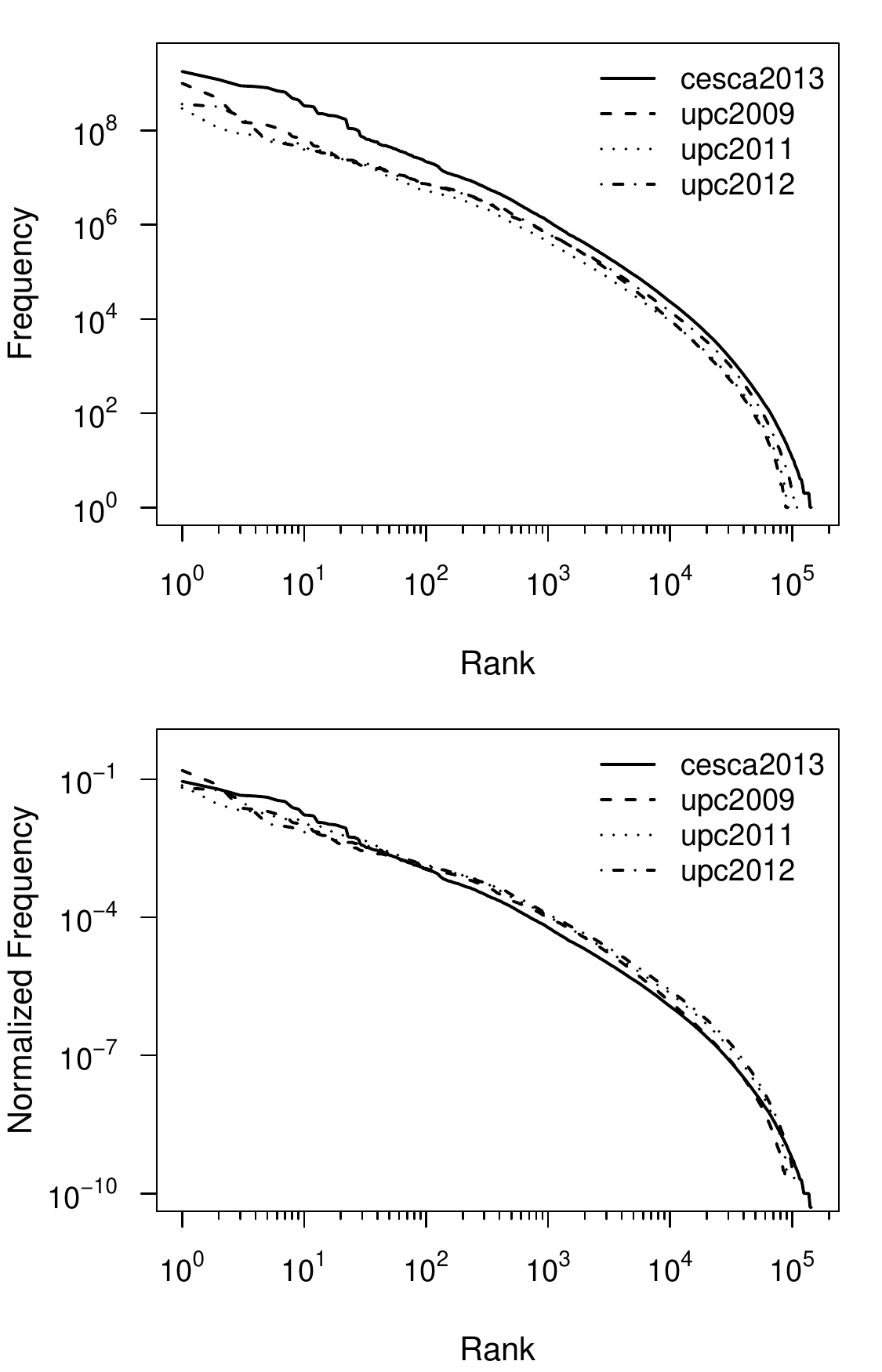}
    \caption{Destination Prefix Popularity}
    \label{fig:pop}
\end{figure}

\begin{figure}[t]
    \centering
    \includegraphics[width=0.5\textwidth, keepaspectratio=true]{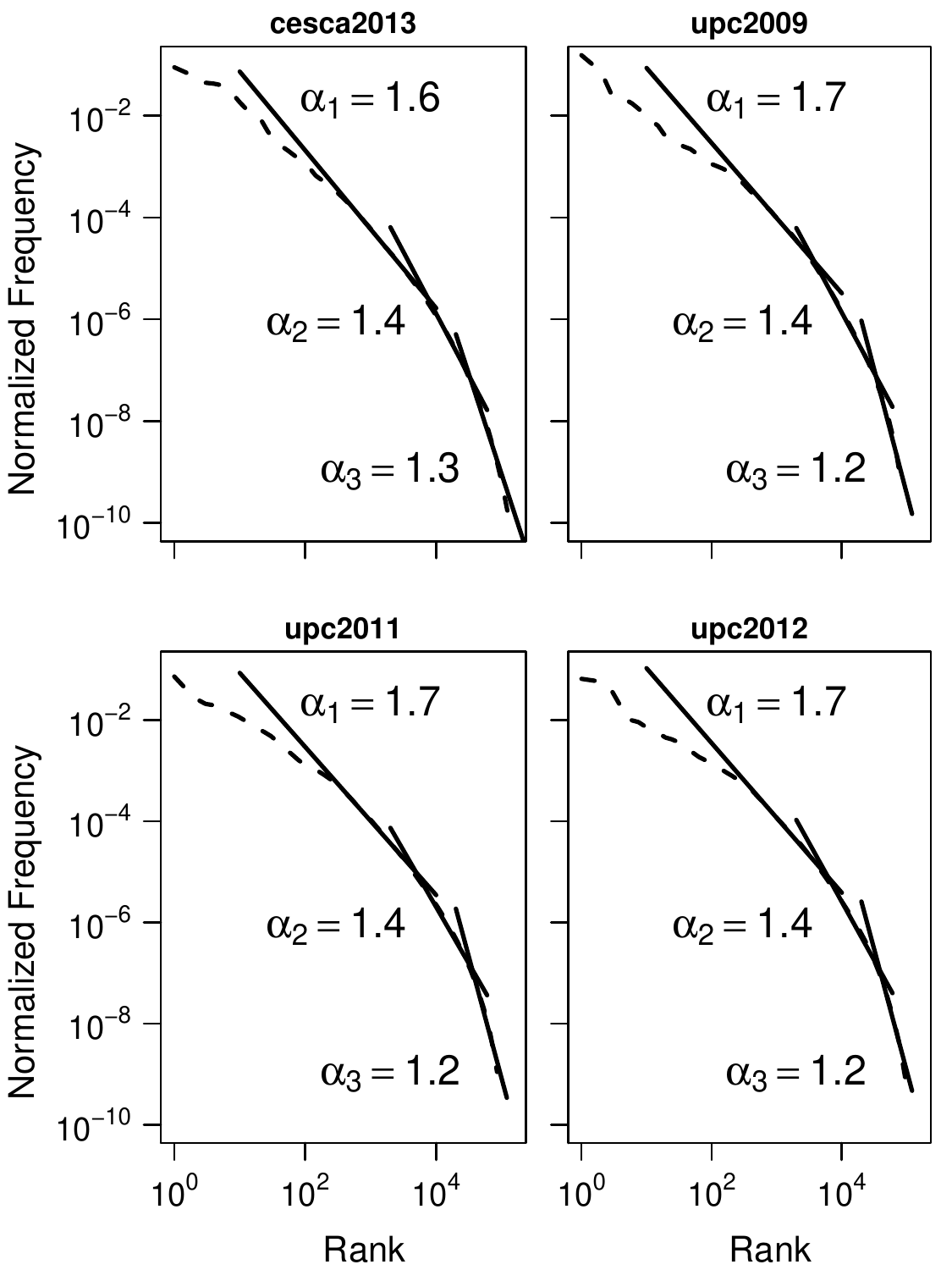}
    \caption{Frequency-rank distribution of destination prefixes and a
    linear least squares fit of the three power-law regiemes.
    $\alpha_i=1+1/s_i$, where $s_i$ is the slope of the $i$th segment.}
    \label{fig:pop_fit}
\end{figure}

Figure~\ref{fig:pop} presents the frequency-rank distributions of our datasets
for both absolute and normalized frequency. A few observations are in
place. First, although clearly not accuretely described by Zipf's law, they
also slightly deviate from a GZipf. Namely, the head of the distribution
presents two power-law regiemes followed by a third that describes the tail as
it can be seen in Fig.~\ref{fig:pop} (down). This may be either
because a one day sample is not enough to obtain accurate statistics in the
Zipf-Mandelbrot head reagion, or because popularity for low ranks follows a
more complex law. Still, we find that for all traces the frequencies of higher
ranks (above 2000) are accurately characterized by two power-law regiemes (see
Fig.~\ref{fig:pop_fit}).

Secondly, the frequency-rank curves for the UPC datasets are remarkably
similar. Despite the  $50\%$ increase of $\texttt{BGP}_{\phi}$ (i.e., $D$),
changes in the Internet content provider infrastructure over a $3.5$ years
period, and perhaps even changes in the local user set, the popularity
distributions are roughly the same. 

Finally, the normalized frequency plots for all traces are similar, in spite
of the large difference in number of packets between CESCA and UPC datasets.
These observations confirm our assumption that growth of the number of users
within the site or of the destination space do not necessarily result in a
change of the popularity distribution.

To confirm that these results are not due to a bias of popularity for larger
prefixes sizes, that is, larger prefixes are more probable to receive larger
volumes of traffic because they contain more hosts, we checked the correlation
between prefix length and frequency. But (not shown here) we didn't find any
evidence in support of this.

\subsection{Prefix Inter-Reference Distance Distribution}

\begin{figure}[t]
    \centering
    \includegraphics[width=0.5\textwidth,keepaspectratio=true]{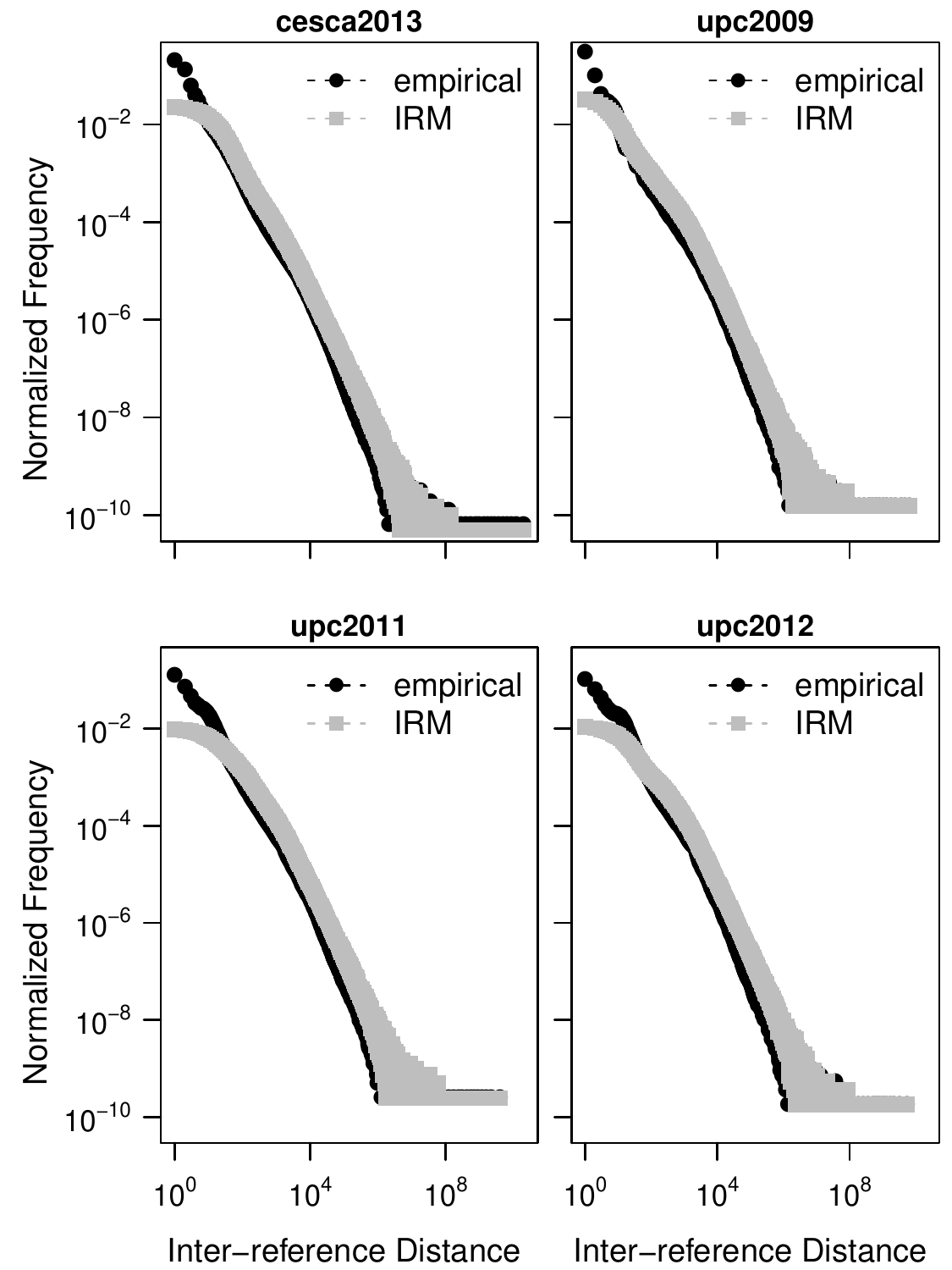}
    \caption{Empirical and IRM generated inter-reference for the four traces}
    \label{fig:ir_comparison}
\end{figure}

We now check if knowledge about the popularity distribution suffices to
accurately characterize the inter-reference distance distribution or if
short-term correlations must also be taken into account. To achieve this, we
use a methodology similar to the one used in~\cite{jin:web_tloc} for Web page
traffic. We first generate random versions of our traces according to the IRM
model, i.e., by considering only the popularity distribution and geometric
inter-reference times, and then compare the resulting inter-reference distance
distributions to the originals. Results are shown in
Fig.~\ref{fig:ir_comparison}. We find that for all traces, popularity alone is
able to account for the greater part of the inter-reference distance
distribution, like in the case of Web requests. The only disagreement is in
the region with distances lower than $100$ where short-term correlations are
important and IRM traces underestimate the probability by a significant
margin.

A rather interesting finding is that the short-term correlations in all traces
are such that the power-law behavior observed for higher distances ($t>100$) is extended
up to distance $1$. In this region, the exact inter-reference distance
equation (\ref{eq:ir_q}) is a poor fit to reality as it follows
the IRM curve. However, the empirical results are apropriately described by our
approximate inter-reference model (\ref{eq:fq_asym}) which avoids IRM's bent by assuming
(\ref{eq:ir_q})'s numerator constant. 

\subsection{Cache Performance}\label{sec:cache_res}

Having found that our assumptions regarding network traffic properties hold in
our datasests we now if the cache model (see~(\ref{eq:msq})
and~(\ref{eq:msr})) is able predict real world LRU cache performance. 

As mentioned in Section~\ref{sec:pop_assumption} and as it may be seen in
Fig.~\ref{fig:pop_fit}, the head of the popularity distribution exhibits two
power-law regiemes instead of one. Then, two options arise, we can either use
the model disregarding the discrepancies or adapt it to consider the low rank
region behavior. For completness, we choose the latter in our evaluation. This
only consists in approximating $p_\nu(\nu)$ (see (\ref{eq:gzipf_pdf})) as
having three regions, each dominated by an exponent $\alpha_i$. Recomputing
(\ref{eq:msr}) we get that the miss rate has three regions, each
characterized by an $\alpha_i$. Choosing the first option would only
result in an overestimation of cache miss rates for low cache sizes.

\begin{figure}[t]
    \centering
    \includegraphics[width=0.5\textwidth, keepaspectratio=true]{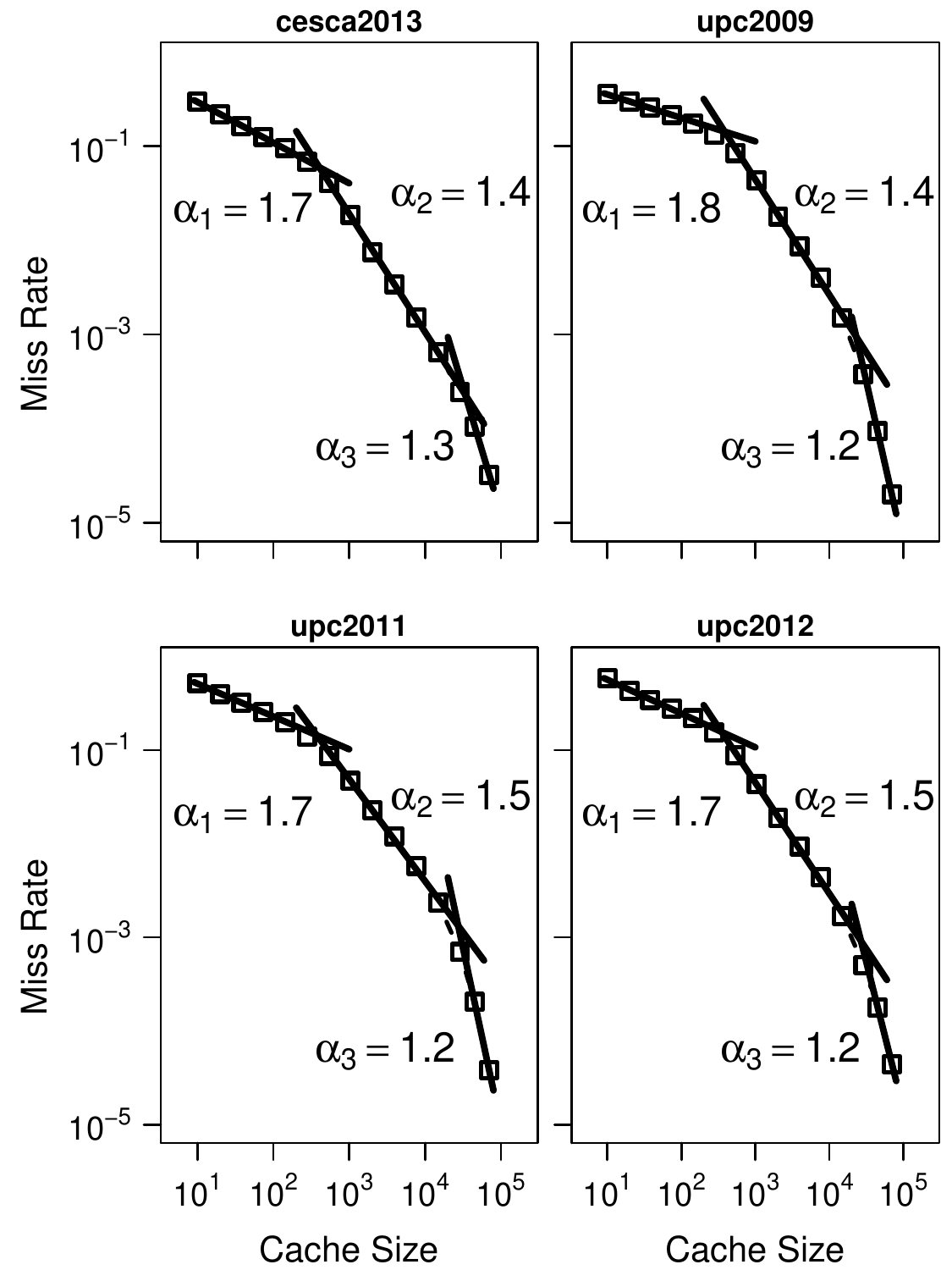}
    \caption{Empirical miss rate with cache size and a linear least-squares
    fit of the exponent for the three power-law regions. Notice the
    similarity with the exponents of the three regions of the popularity
    distribution in Fig~\ref{fig:pop_fit}. }
    \label{fig:mr_fit}
\end{figure}

\begin{figure}[t]
    \centering
    \includegraphics[width=0.5\textwidth,keepaspectratio=true]{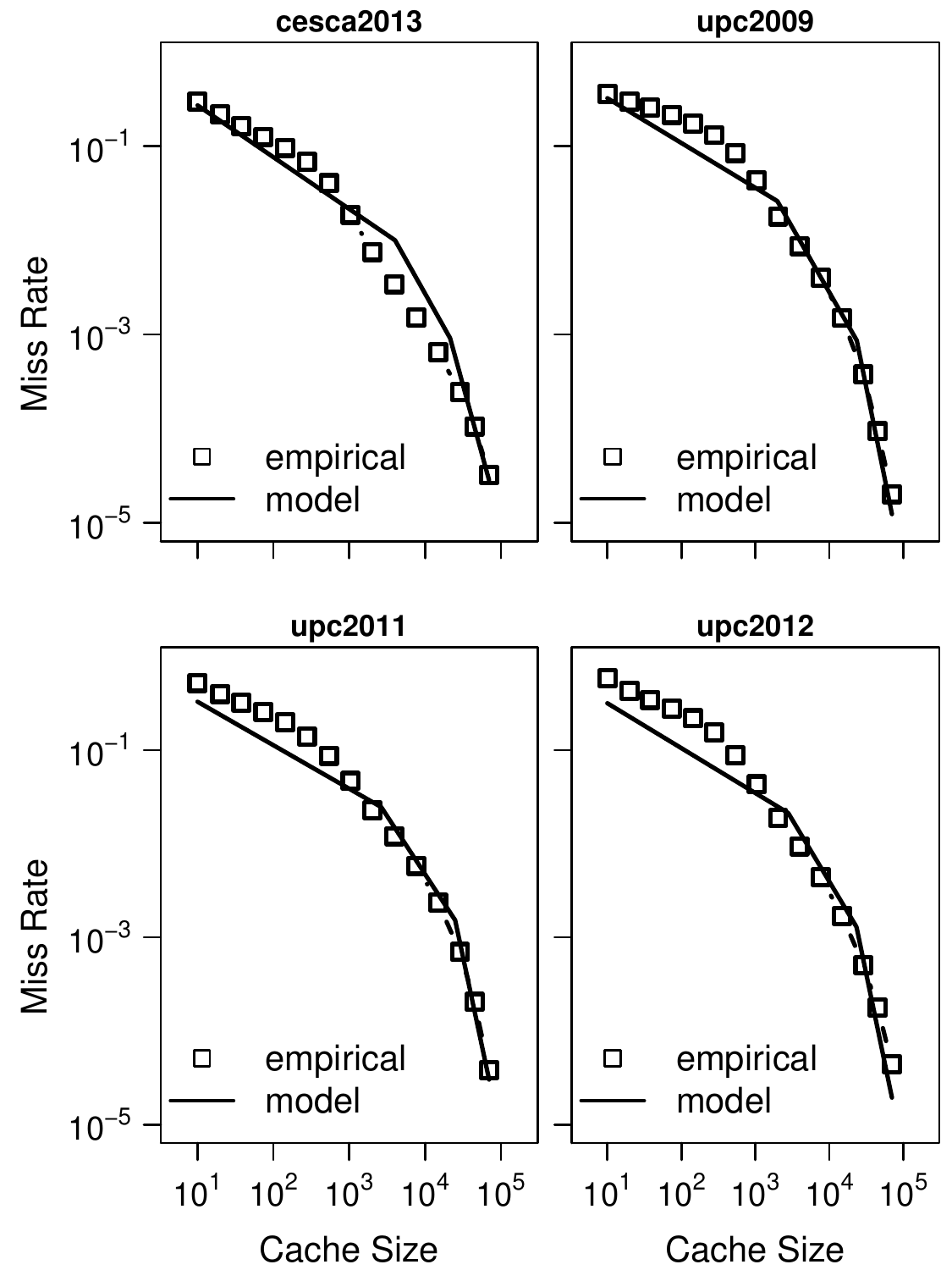}
    \caption{Empirical miss rate with cache size together with a fit by (\ref{eq:msq}) and
    (\ref{eq:msr})}
    \label{fig:mr_model}
\end{figure}

To contrast the model with the empirical observations, we performed a linear
least squares fit of the three regions of the popularity distribution. This
allowed us to determine the exponents $\alpha_i$, computed as
$1 + 1/s_i$ where $s_i$ is the slope of the $i$th segment,  and to
roughly approximate the frequencies $\nu_{k1}$ and $\nu_{k2}$ at which the
segments intersect. Using them as input to (\ref{eq:msq}) we get a cache miss
rate estimate as shown in Fig.~\ref{fig:mr_model}. Generally, we see that the
model is a remarkably good fit for the large cache sizes but constantly underestimates the
miss rate for sizes lower than 1000. This may be due to the
poor fit of the popularity for low ranks. Nevertheless a more elaborate fitting of
$\nu_{k1}$ and $\nu_{k2}$ should provide better results as it may be seen
in Fig.~\ref{fig:mr_fit} where we performed a linear least squares fit of the three
power law regions of the cache miss rate. Knowing that the slope of the cache
miss rate is $s_i=1-1/(\alpha_i-1)$ (see (\ref{eq:mtq})), we computed the
exponents as depicted in the figure. Comparison with those computed in
Fig.~\ref{fig:pop_fit} shows they are very similar. Overall, we can conclude
that the model accurately predicts cache performance. 

\section{Related Work}\label{sec:rw}
Denning was first to recognize the phenomenon of temporal locality in his
definition of the working-set~\cite{denning:ws_model} and together with
Schwartz established the fundamental properties that characterize
it~\cite{denning:ws_properties}. Although initially designed for the analysis
of page caching in operating systems, the ideas were later reused in other
fields including Web page and route caching.

In~\cite{breslau:web_and_zipf} Breslau et al. argued that empirical evidence
indicates that Web requests popularity distribution is Zipf-like of exponent
$\alpha < 1$. Using this finding and the assumption that temporal locality is
mainly induced through long-term popularity, they showed that the asymptotic
miss rates of an LFU cache, as a function of the cache size, is a power law of
exponent $1-\alpha$. In this paper we argue that GZipf with exponents greater
than $1$ is a closer fit to real popularity distributions and obtain a more
general LRU cache model. We further use the model to determine the scaling
properties of the cache. 

Jin and Bestavros showed in~\cite{jin:web_tloc} that the inter-reference
distribution is mainly determined by the the long-term popularity and only
marginally by short-term correlations. They also proved that the
inter-reference distribution of a reference string with Zipf-like popularity
distribution is proportional to $1/t^{2-1/\alpha}$. We build upon their work
but also extend their results by both considering a GZipf popularity
distribution and by using them to deduce an LRU cache model. 

In the field of route caching, Feldmeier~\cite{feldmeier:rt_cache} and
Jain~\cite{jain:dst_locality} were among the first to evaluate the possibility
of performing destination address caching by leveraging the locality of
traffic in network environments. Feldmeier found that locality could be
exploited to reduce routing table lookup times on a gateway router while Jain,
discovered that deterministic protocol behavior limits the benefits of
locality for small caches. The works, though fundamental, bear no practical
relevance today as they were carried two decades ago, a time when the Internet
was still in its infancy. 

Recently, Kim et al.~\cite{kim:rcaching} performed a measurement study within
the operational confinement of an ISP's network and showed the feasibility of
route caching. They show by means of an experimental evaluation that LRU cache
eviction policy performs close to optimal and better than LFU. Also, they
found  that prefix popularity distribution is very skewed and that working-set
size is generally stable with time. These are in line with our empirical
findings and provide practical confirmation for our assumption that the
popularity distribution can be described as a GZipf. 

Several works have previously looked at cache performance in loc/id split
scenarios considering LISP as a reference implementation. Iannone et
al.~\cite{iannone:lcache} performed an initial trace driven study of the LISP
map-cache performance while Kim et al.~\cite{jkim:lcache} have both extended
and confirmed the previous results with the help of a larger, ISP trace.
Zhang et al.~\cite{zhang:lcache} performed a trace based Loc/ID mapping cache
performance analysis assuming a LRU eviction policy and using traffic captured
at two egressing links of the China Education and Research Network backbone
network. Although methodologies differ between the different papers, in all
cases the observed LISP cache miss rates were found to be relatively small.
This, again, indirectly confirms the skewness of the popularity distribution
and its stability at least for short time scales. 

Finally, in~\cite{coras:lcache_n} we devised an analytical model for the LISP
cache size starting from empirical average working-set curves, using the
working-set theory. Our goal was to model the influence of locality on cache
miss rates whereas here, we look to understand how cache performance scales with
respect to defining parameters, that is, the popularity distribution, the size
of the LISP site and the size of the EID space, of network traffic.

\section{Conclusions}\label{sec:conclusions}

LISP offers a viable solution to scaling the core routing infrastructure of
the Internet by means of a location/identity split. However this forces edge
domain routers to cache location to identity bindings for timely operations.
In this paper we answer the following question: does the newly introduced LISP
edge cache scale?

Our findings show that the miss rate scales constantly O(1) with the number of
users as well as with the number of destinations. For this, we start from two
assumptions: (i) the popularity of destination prefixes is described by a
GZipf distribution and (ii) temporal locality is predominantly determined by
long-term popularity. Fundamentally, these assumptions are often observed to
hold in the Internet~\cite{sarrar:leverage_zipf, kim:rcaching} but also in
other fields such as web traffic~\cite{breslau:web_and_zipf}, on-demand
video~\cite{cha:itube} or even linguistics~\cite{zipf:principle_least_effort}.
Arguably, they are inherent to human nature and, as such, are expected to hold
in the foreseeable future. Nevertheless, in the paper we also show that if the
converse holds, then cache size scales linearly O(N) with the number of
destinations.

At the time of this writing there is an open debate on how the Internet should
look like in the near future and in this context, it is important to analyze
the scalability of the various future Internet architecture proposals. This
paper fills this gap, particularly for the Locator/ID split architecture.
Furthermore, our results show that edge networks willing to deploy LISP will
not face scalability issues -as long as both assumptions hold- in the size of
their map-cache, even if the edge network itself becomes larger (i.e., more
users) or the Internet grows (i.e., more prefixes).
 
\section*{Acknowledgements} 
The authors would like to express their gratitude to Damien Saucez and Chadi
Barakat for their insightful comments. This work has been partially supported
by the Spanish Ministry of Education under scholarship AP2009-3790, research
project TEC2011-29700-C02, Catalan Government under project 2009SGR-1140 and a
Cisco URP Grant.

\small

\normalsize
\end{document}